**Specification of an extensible and portable file format for electronic structure and crystallographic data**.

X. Gonze (1,2,+), C.-O. Almbladh (1,3), A. Cucca (1,4), D. Caliste (1,2,5), C. Freysoldt (1,6),
M. A. L. Marques (1,7,8), V. Olevano (1,4,9), Y. Pouillon (1,2,10), M.J. Verstraete (1,11)

(1) European Theoretical Spectroscopy Facility (ETSF)
(2) Université Catholique de Louvain, Louvain-la-Neuve (Belgium)
(3) University of Lund, Lund (Sweden)
(4) LSI, CNRS-CEA, Ecole Polytechnique, Palaiseau (France)
(5) C.E.A. Grenoble, Grenoble (France)
(6) Fritz-Haber-Institut, Berlin (Germany)
(7) U. Lyon 1, Villeurbanne (France)
(8) U. Coimbra, Coimbra (Portugal)
(9) Institut NEEL, CNRS and U. Joseph Fourier, Grenoble (France)
(10) Universidad del Pais Vasco UPV/EHU, Donostia-San Sebastiàn (Spain)
(11) U. York, York (United Kingdom)

+ Corresponding author.

**Abstract**

In order to allow different software applications, in constant evolution, to interact and exchange data, flexible file formats are needed. A file format specification for different types of content has been elaborated to allow communication of data for the software developed within the European Network of Excellence "NANOQUANTA", focusing on first-principles calculations of materials and nanosystems. It might be used by other software as well, and is described here in detail. The format relies on the NetCDF binary input/output library, already used in many different scientific communities, that provides flexibility as well as portability accross languages and platforms. Thanks to NetCDF, the content can be accessed by keywords, ensuring the file format is extensible and backward compatible.

**Introduction**

Standardization of file formats is a prerequisite for seamless data exchange between applications. Different standards might be used concurrently, as, for example, JPEG, TIFF, PICT, PNG, GIF, EPS, ... , that encode 2D images. The development of associated conversion software quickly follows the definition of the specification.

In the scientific community developing software for first-principles calculations of material properties, based on electronic structure, the primary concern of developers (usually physicists and not software engineers) has always been the improvement of capabilities, usually leaving aside exchange of data. Typically, the existing file formats match the software capabilities at one moment in time, without much care about file standardization. The default policy adopted for large files is to use the binary representation provided by the native programming language (e.g. FORTRAN or C/C++), with several drawbacks: (i) lack of portability between big-endian and little-endian platforms (and vice-versa), or between 32-bit and 64-bit platforms; (ii) difficulties to read the files written by F77/90 codes from C/C++ software (and vice-versa); (iii) lack of extensibility, as one file produced for one version of the software might not be readable by a past/forthcoming version.

Software scientists and several other scientific communities addressed such problems already a long time ago. In particular, the meteorology/climatology community developed the NetCDF (Network Common Data Form) library [1,2], that provides tools for generating/reading files that are (1) portable accross platforms (e.g. independent of the native binary encoding -big/little endian), (2) independent of the programming language (C/C++, F77/90, Java, Perl, Python, ...), (3) as economical as binary files concerning the storage space, and (4) addressed by keywords, usually solving the problem of backward compatibility. This appears in retrospect as a crucial step in the cross-validation of simulations performed by different groups, working with different platforms and codes.

The idea of standardization of file formats is not new in the electronic-structure community [3]. However, it proved difficult to achieve without a formal organization, gathering code developers involved in different software project, with a sufficient incentive to realize effective file exchange between such software. In the EU Nanoquanta Network of Excellence and the recently started ``European Theoretical Spectroscopy Facility'', the standardization of file formats is an important step towards integration. For sake of brevity, we will refer to these two projects as NQ/ETSF. Discussions within NQ/ETSF have lead to a specification document [4], a large part of which will be presented here. This specification has already been implemented in different software, either on the basis of a full-fledged library, ETSF_IO [5,6] or using an early exploratory interface [7]. Additional information can be found in Ref. [8].

Actually, NetCDF is not the only possible choice of library in the above-mentioned context. HDF [9] also implements a software solution that addresses the portability and extensibility problems, with economical use of storage space. NetCDF is simpler to use if the data formats are flat, while HDF has definite advantages if one is dealing with hierarchical formats. Typically, we will need to describe many different multi-dimensional arrays of real or complex numbers, for which NetCDF is an adequate tool. Although a data specification might be presented irrespective of the library actually used for the implementation, such a freedom might lead to implementations using incompatible file formats (like NetCDF and XML for instance). This possibility would in effect block the expected standardization gain. Thus, as a part of the standardization, we require the future implementations of our specification to rely on the NetCDF library only.

The NQ/ETSF specification relying on NetCDF focuses on the data that needs somewhat large storage space (representation of wavefunctions and densities/potentials, and the additional auxiliary information). In the electronic-structure community, one needs also to exchange atomic data (e.g. pseudopotentials or data for the projector-augmented wave methodology), that consumes much less storage space. The compact representation brought by NetCDF can be by-passed in favour of text encoding. We are aware of several standardization efforts [10,11], relying on XML, which emphasize addressing by content to represent such atomic data.

Section 1 presents a brief account of the major characteristics of NetCDF, then outlines our generic design decisions for the NQ/ETSF file format specification. In section 2, we list the few global and generic attributes, dimensions, optional variables, and naming conventions of NQ/ETSF NetCDF files. In sections 1 and 2, the reader will not find a full description of the use of NetCDF, but only sufficient NetCDF background to understand our design decisions and the NQ/ETSF specifications themselves. In sections 3, 4 and 5, we describe the specification for the crystallographic data, density/potentials and wavefunction parts of the NQ/ETSF files, leaving aside several technical elaborations, especially those concerning the distribution of the data in different files. Section 6 concludes and briefly presents the library associated to the specification, and the current use of the specification.

## 1. Guidelines

In a NetCDF file, a series of numerical arrays can be stored, each under the name of a variable, that might possess one or several attributes and possibly one or several dimensions. Global attributes of the file (not associated to one particular variable) can also be stored.

The NetCDF library [1,2] provides functions to initialize a NetCDF file, to create variable names and their dimensions in this file, to associate attributes to them, to define their dimensions, and to store the associated numerical data. It provides also functions to inquire about the content of a file (names of variables, associated dimensions and attributes), to access the information associated to a variable name (in full or by segments), to copy it, to rename attributes or variables, or to delete some of its content. Thanks to the powerful inquiry functions, it is possible to get full information from a NetCDF file without *a priori* knowledge of its content. In most cases, however, the user knows the name of the variables, the content he/she would like to retrieve, and the associated dimensions. The ability of NetCDF to retrieve the information, irrespective of the actual physical layout of the file, is a key characteristic allowing exchange of data between different software (and also different versions of the same software), that contrasts with the rigidity of the usual binary representations.

The major step in the specification of NQ/ETSF files based on NetCDF consists in the definition of variable names, associated dimensions and attributes, and accurate definition of the corresponding physical quantities. Our first design decision, stemming from the flexibility of NetCDF, dealt with the different types of variables/attributes and associated numerical information. We distinguished four different types of variables:

(A) The actual numerical data (which defines whether a file contains wavefunctions, a density, etc), for which a name must have been agreed in the specification.

(B) The auxiliary data that is mandatory to make proper usage of the actual numerical data of A-type. The name and description of this auxiliary information is also agreed.

(C) The auxiliary data that is not mandatory to make proper usage of the A-type numerical data, but for which a name and description has been agreed in the specification.

(D) Other data, typically code-dependent, whose availability might help the use of the file for a specific code. The name of these variables should be different from the names chosen for agreed variables of A-C types. Such type D data might even be redundant with type A-C data.

The four types are compatible with a file being sufficiently complete for use by many different codes, though adapted to the specific usage by each of them. The NQ/ETSF file specification is based on this generic classification: for three sets of numerical data (A-type : density/potential ; crystallographic ; wavefunctions), we define the auxiliary data that is mandatory to make proper usage of it (B-type variables). In addition, we provide names for variables that can be either mandatory or not (in the context of a file containing a density/potential, or a wavefunction, or crystallographic data, or other large numerical data not yet taken into account), but for which a NetCDF description has been agreed.

Some technical details concerning the use of NetCDF files will apply to all specifications in the NQ/ETSF framework:
(1) Concerning the variable names, long names should be chosen, as close as possible to natural language (so inherently self-descriptive, e.g. "number_of_atoms").
(2) All variable names are lower case, except the global attribute "Conventions" - a name agreed by the NetCDF community.
(3) Underscores are used to replace blanks separating words.
(4) In the specification of the dimensions of the variables, we used row-major storage for multi-dimensional arrays (the fast indices are right-most). FORTRAN uses column-major storage (the fast indices are left-most), so that the order of indices has to be reversed in FORTRAN.

Even if the chosen names are close to natural language, the concepts that are embodied in the variables stored in the NQ/ETSF files are inherently complex. The specification presently described includes for each name a brief description, but mostly focuses on the information needed to specify the content and structure of a file. In order to complete the present specification with a thorough description of each theoretical concept, we refer to the book by R. Martin [12]. This recently published book contains numerous references, and covers all the Density Functional Theory background on which we based the major part of the present specification. Some data needed for more advanced computations are related to the GW approximation, for which we refer to the review presentation Ref. [13].

## 2. General specifications for NQ/ETSF NetCDF files

### 2.1. Global attributes of NQ/ETSF NetCDF files

Global attributes are used for a general description of the file, mainly the file format convention. Important data is not contained in attributes, but rather in variables.

Table 1 gather specifications for required attributes in any NQ NetCDF files. Table 2 presents optional attributes for NQ/ETSF NetCDF files.

*Detailed description* (tables 1 and 2)

**file_format** Name of the file format for NQ/ETSF wavefunctions.
**file_format_version** Real version number for file format (e.g. 2.2 ).
**Conventions** NetCDF recommended attribute specifying where the conventions for the file can be found on the Internet.

**history** NetCDF recommended attribute : each code modifying/writing this file is encouraged to add a line about itself in the history attribute. char(1024) allows for 12 additions of at most 80 characters.
**title** Short description of the content (system) of the file.

### 2.2. Generic attributes of variables in NQ/ETSF NetCDF files

A few attributes might apply to a large number of variables. They are gathered in Table 3 .

*Detailed description* (table 3)

**units** It is one of the NetCDF recommended attributes, but it only applies to a few variables in our case, since most are dimensionless. For dimensional variables, it is required. The use of atomic units (corresponding to the string "atomic units") is advised throughout for portability. If other units are used, the definition of an appropriate scaling factor to atomic units is mandatory. Actually, the definition of the name "units" in the NQ/ETSF files is only informative : the "scale_to_atomic_units" information should be the only one used to read the file by machines.

**scale_to_atomic_units** If "units" is something other than the character string "atomic units" (based on Hartree for energies, Bohr for lengths) one requests the definition of an appropriate scaling factor. The appropriate value in atomic units is obtained by **multiplying** the number found in the variable by the scaling factor. Examples:
units="eV" => scale_to_atomic_units = 0.036749326
units="angstrom" => scale_to_atomic_units = 1.8897261
units="parsec" => scale_to_atomic_units = 5.8310856e+26
This can be used to deal with unknown units.
Note that the recommended values for the fundamental constants is available on the Web [14].

## 2.3. Flag-like attributes

"Flag-like" attributes can take the values "yes" and "no". When such attributes are written, they should be written in full length and small letters. When they are read, only the first character needs to be checked.

## 2.4. Dimensions

Dimensions are used for one- or multidimensional variables. The NetCDF interface adapts the dimension ordering to the programming language used. The notation here is C-like, i.e. row-major storage, the last index varying the fastest. In FORTRAN, a similar memory mapping is obtained by reversing the order of the indices. When implementing new reading interfaces, the dimension names can be used to check the dimension ordering. The dimension names help also to identify the meaning of certain dimensions in cases where the number alone is not sufficient.

The list of variables that specify dimensions in NQ NetCDF files is given in Table 4 and 5. Table 4 list the dimensions that are not supposed to lead to a splitting, see Ref.[4], while table 5 lists the dimensions that might be used to define a splitting (e.g. in case of distributed data).

*Detailed description* (table 4)

**character_string_length** The maximum length of string variables (attributes may be longer).

**real_or_complex_coefficients** Needed to specify whether the variable coefficients_of_wavefunctions (table 14) is real or complex

**real_or_complex_density** Needed to specify whether the variable density (table 10) is real or complex

**real_or_complex_gw_corrections** Needed to specify whether the variable gw_corrections (table 16) is real or complex

**real_or_complex_potential** Needed to specify whether the variables exchange_potential, correlation_potential, and exchange_correlation_potential (table 11) are real or complex

**real_or_complex_wavefunctions** Needed to specify whether the variable real_space_wavefunctions (table 15) is real or complex .

**number_of_cartesian_directions** Used for absolute coordinates.

**number_of_reduced_dimensions** Used for reduced (also called relative) coordinates in reciprocal or real space.

**number_of_vectors** Used to distinguish the vectors when defining their relative/reduced coordinates.

**number_of_symmetry_operations** The number of symmetry operations.

**number_of_atoms** The number of atoms in the unit cell.

**number_of_atom_species** The number of different atom species in the unit cell.

**symbol_length** Maximum number of characters for the chemical symbols

*Detailed description* (Table 5)

**max_number_of_states** The maximum number of states

**number_of_kpoints** The number of kpoints

**number_of_spins** Used to distinguish collinear spin-up and spin-down components :

      1 for non-spin-polarized or spinor wavefunctions

      2 for collinear spin (spin-up and spin-down)

**number_of_spinor_components** For non-spinor wavefunctions, this dimension must be present and equal to 1. For spinor wavefunctions this dimension must equal to 2.

**number_of_components** Used for the spin components of spin-density matrices :

      1 for non-spin-polarized

      2 for collinear spin (spin-up and spin-down)

      4 for non-collinear spin (average density, then magnetization vector in

            cartesian coordinates x,y and z)

**max_number_of_coefficients** The (maximum) number of coefficients for the basis functions at each k-point, except in the case of real space grids (see next lines)

**number_of_grid_points_vector1** The number of grid points along direction 1 in the unit cell in real space, for dimensioning the wavefunction coefficients (an alternative to max_number_of_coefficients)

**number_of_grid_points_vector2** Same as number_of_grid_points_vector1, for the second direction.

**number_of_grid_points_vector3** Same as number_of_grid_points_vector1, for the third direction.

To clarify the interplay between number_of_spins, number_of_components, and number_of_spinor_components, note the different following magnetic or non-magnetic cases:

Non-spin-polarized :

number_of_spins=1 , number_of_spinor_components=1, number_of_components=1

Collinear spin-polarized :

number_of_spins=2, number_of_spinor_components=1, number_of_components=2

Non-collinear spin-polarized :

number_of_spins=1, number_of_spinor_components=2, number_of_components=4

## 2.5. Optional variables

In order to avoid the divergence of the formats in the additional data, we propose names and formats for some information that is likely to be written to the files. None of these data is mandatory for the file formats to be described later. Some of the proposed variables contain redundant information.

Tables 6 to 8 present these optional variables, grouped with respect to their physical relevance: atomic information, electronic structure, and reciprocal space.

*Detailed description* (tables 7 to 10)

**valence_charges** Ionic charges for each atom species.

**pseudopotential_types** Type of pseudopotential scheme

      = "bachelet-hamann-schlueter", "troullier-martins", "hamann",

        "hartwigsen-goedecker-hutter", "goedecker-teter-hutter" ...

**number_of_electrons** Number of electrons in the elementary cell.

**exchange_functional** String describing the functional used for exchange: names should be taken from the Nanoquanta XC library specifications (see Ref.[4]).

**correlation_functional** String describing the functional used for correlation: Lee Yang Parr or Colle-Salvetti etc... names should be taken from the Nanoquanta XC library specifications (see Ref.[4]).

**fermi_energy** Fermi energy corresponding to occupation numbers.

**smearing_scheme** Smearing scheme used for metallic or finite temperature occupation numbers = "gaussian", "fermi-dirac", "cold-smearing", "methfessel-paxton-n" for n=1 ... 10

**smearing_width** Smearing width used with scheme above.

**kinetic_energy_cutoff** Cutoff used to generate the plane-wave basis set.

**kpoint_grid_vectors** Basis vectors for kpoint grid if it is homogeneous. Given in the coordinates of reciprocal space primitive vectors.

**kpoint_grid_shift** Shift for offset of grid of kpoints. Used with both kpoint_grid_vectors and monkhorst_pack_folding.

**monkhorst_pack_folding** This indicates the ``folding'' for regular kpoint grids (e.g. Monkhorst-Pack Phys. Rev. B 13, 5188 (1976)). An alternative to kpoint_grid_vectors.

2.6 Naming conventions

NetCDF files, that respect the NQ/ETSF specifications described in the present document, should be easily recognized, thanks to the final substring "-etsf.nc" . The appendix ".nc" is a standard convention for naming NetCDF files [2].

**3. Specification for files containing crystallographic data**

A NQ/ETSF NetCDF file for crystallographic data should contain the following set of mandatory information :
(1) The three attributes defined in Table 1
(2) The following dimensions from Table 4 (dimensions that do not lead to a splitting) :
     - number_of_cartesian_directions
     - number_of_vectors
     - number_of_atoms
     - number_of_atom_species
     - number_of_symmetry_operations
(3) The following variables defined in Table 9 :
     - primitive_vectors
     - reduced_symmetry_matrices
     - reduced_symmetry_translations
     - space_group

- atom_species
- reduced_atom_positions

(4) At least one of the following variables defined in Table 9, to specify the kind of atoms :
- atomic_numbers
- atom_species_names
- chemical_symbols

The use of "atomic_numbers" is preferred. If not available, "atom_species_names" will be preferred over "chemical_symbols". In case more than one such variables are present in a file, the same order of preference should be followed by the reading program.

*Detailed description* (table 9)

**primitive_vectors**  The primitive vectors, expressed in cartesian coordinates.

Symmetry operations are defined in real space, with reduced coordinates.
A symmetry operation in real space sends the input point r to the output point r', with

$$r\,'^{red}_{\alpha} = \sum_{\beta} S^{red}_{\alpha\beta} r^{red}_{\beta} + t^{red}_{\beta}$$

The matrix S, in reduced coordinates, is contained in the array **reduced_symmetry_matrices** of Table 9, while the vector t, in reduced coordinates, is contained in the array **reduced_symmetry_translations** of the same Table. There might be a confusion between the two dimensions "number_of_reduced_dimensions" of this variable. In the C ordering, the last one corresponds to the $\beta$ index in the above-mentioned formula.

The first symmetry operation must always be unity with translation vector (0,0,0). If all translations are zero, the attribute **symmorphic** for reduced_symmetry_matrices should be set to "yes".

**space_group**  Space group number according to international tables of crystallography (from 1 to 232)

**atom_species**  Types of each atom in the unit cell. Note that the first type of atom has number "1", and the last type of atom has number "number_of_atom_species".

**reduced_atom_positions**  Positions of the different atoms in the unit cell in relative/reduced coordinates.

**atomic_numbers**  Atomic number for each atom species. If it does not refer to an "usual" atom (e.g. fractional charge atoms or similar), a non-integer number or zero may be used, but it is strongly adviced then to also specify the atom_species_names variable.

**atom_species_names**  Descriptive name for each atom species = "H" "Ga" plus variants like "Ga-semicore" "C-1s-corehole" "C-sp2" "C1"

**chemical_symbols**  Chemical symbol for each atom species (as in periodic table). If not appropriate (fractional charge atoms or similar), "X" may be used.

**symmorphic** Flag-type attribute (see section 2.3), needed for the variables reduced_symmetry_matrices and reduced_symmetry_translations.

**4. Specification for files containing a density and/or a potential**

A NQ/ETSF NetCDF file for a density should contain the following set of mandatory information :
(1) The three attributes defined in Table 1
(2) The following dimensions from Table 4 :
      - number_of_cartesian_directions
      - number_of_vectors
      - real_or_complex_density and/or real_or_complex_potential
(3) The following dimensions from Table 5 :
      - number_of_components
      - number_of_grid_points_vector1
      - number_of_grid_points_vector2
      - number_of_grid_points_vector3
(4) The primitive vectors of the cell, as defined in Table 9
(5) The density or potential, as defined in Table 10 or 11.

A NQ/ETSF NetCDF exchange, correlation, or exchange-correlation potential file should contain at least one variable among the three presented in Table 11 in replacement of the specification of the density. The type and size of such variables are similar to the one of the density. The other variables required for a density are also required for a potential file. Additional NQ or software-specific information might be added, as described previously.

A density in such a format (represented on a 3D homogeneous grid) is suited for the representation of smooth densities, as obtained naturally from norm-conserving pseudopotential calculations using plane waves. The same apply to potentials.

**5. Specification for files containing the wavefunctions**

5.1. Specification

A NQ/ETSF NetCDF file "containing the wavefunctions" should contain at least the information needed to build the density from this file. Also, since the eigenvalues are intimately linked to eigenfunctions, it is expected that such a file contain eigenvalues. Of course, files might contain less information than the one required, but still follow the naming convention of NQ/ETSF. It might also contain more information, of the kind specified in other tables of the present document.

A NQ/ETSF NetCDF file "containing the wavefunctions" should contain the following set of mandatory information :
(1) The three attributes defined in Table 1
(2) The following dimensions from Table 4 (dimensions that do not lead to a splitting) :
      - character_string_length
      - number_of_cartesian_directions
      - number_of_vectors
      - real_or_complex_coefficients and/or real_or_complex_wavefunctions
      - number_of_spinor_components
      - number_of_symmetry_operations

       - number_of_reduced_dimensions

(3) The following dimensions from Table 5 (dimensions that might lead to a splitting) :

       - max_number_of_states

       - number_of_kpoints

       - number_of_spins

(4) In case of a real-space wavefunctions, the following dimensions from Table 5 :

       - number_of_grid_points_vector1

       - number_of_grid_points_vector2

       - number_of_grid_points_vector3

       or, in case of a wavefunction given in terms of a basis set, the following dimensions from Table 5:

       - max_number_of_coefficients

(5) The primitive vectors of the cell, as defined in Table 9 (variable primitive_vectors)

(6) The symmetry operations, as defined in Table 9 (variables reduced_symmetry_translations, reduced_symmetry_matrices)

(7) The information related to each kpoint, as defined in Table 12

(8) The information related to each state (including eigenenergies and occupation numbers), as defined in Table 13

(9) In case of basis set representation, the information related to the basis set, and the variable coefficients_of_wavefunctions , as defined in Table 14

(10) In case of real-space representation, the variable real_space_wavefunctions, see Table 15.

*Detailed description* (Table 12)

**reduced_coordinates_of_kpoints** k-point in relative/reduced coordinates

**kpoint_weights** k-point integration weights. The weights must sum to 1. See the description of the density construction, section 5.2.

*Detailed description* (Table 13)

**number_of_states** Number of states for each kpoint, if varying (the attribute **k_dependent** must be set to yes). Otherwise (the attribute **k_dependent** must be set to no), might not contain any information, the actual number of states being set to max_number_of_states.

**eigenvalues** One-particle eigenvalues/eigenenergies. Should be 0 if unknown.

**occupations** Occupation numbers. Full occupation for spin-unpolarized cases (number_of_spins = 1 AND number_of_spinor_components = 1) is 2, otherwise it is 1. See section 5.2 .

**k_dependent** Flag-type attribute (see section 2.3), needed for the variables number_of_states, number_of_coefficients, and reduced_coordinates_of_plane_waves.

*Detailed description* (Table 14)

**basis_set** Type of basis set used if not in a real-space grid.

**number_of_coefficients** Number of basis function coefficients for each kpoint, if varying (the attribute **k_dependent** must be set to yes). Otherwise (the attribute **k_dependent** must be set to no), might not contain any information, the actual number of coefficients being set to max_number_of_coefficients.

**reduced_coordinates_of_plane_waves** Plane-wave G-vectors in relative/reduced coordinates. If the attribute **k_dependent** is set to no, then the dimension [number_of_kpoints] must be omitted. If the attribute **used_time_reversal_at_gamma** is set to yes (only allowed for the plane wave basis set), then, for the Gamma k point - reduced_coordinates_of_kpoints being equal to (0 0 0) - the time reversal symmetry has been used to nearly halve the number of plane waves, with the coefficients of the wavefunction for a particular reciprocal vector being the complex conjugate of the coefficients of the wavefunction at minus this reciprocal vector. So, apart the origin, the coefficient of only one out of each pair of corresponding plane waves ought to be specified. Note also that the dimension **max_number_of_coefficients** actually governs the size of **reduced_coordinates_of_plane_waves,** so only when the gamma kpoint is present alone, will the size of the file effectively be reduced by the factor of two.

**coefficients_of_wavefunctions** Wavefunction coefficients. The wavefunctions must be normalized to 1, i.e. the sum of the absolute square of the coefficients of one wavefunction must be 1. See section 5.2 The attribute **used_time_reversal_at_gamma** must be used in the same way as for the variable **reduced_coordinates_of_plane_waves** .

**used_time_reversal_at_gamma** Flag-type attribute (see section 2.3), that can be used for the variables **reduced_coordinates_of_plane_waves** and **coefficients_of_wavefunctions**

*Detailed description* (Table 15)

**real_space_wavefunctions** Wavefunction coefficients. Unlike for explicit basis set, the wavefunctions must be normalized to 1 per unit cell, i.e. the sum of the absolute square of the coefficients of one wavefunction, for all points in the grid, <u>divided by the number of points</u> must be 1. See section 5.2 . Note that this array has a number of dimensions that exceeds the maximum allowed in FORTRAN (that is, seven). This leads to practical problems only if the software to read/write this array attempts to read/write it in one shot. Our suggestion is instead to read/write sequentially parts of this array, e.g. to write the spin up part of it, and then, add the spin down. This might be done using Fortran arrays with at most seven dimensions.

*Detailed description* (Table 16)

**gw_corrections** GW-corrections to one-particle eigenvalues (see Table 13). Imaginary part (originating from the non-hermiticity) is optional. Should be 0 if unknown.

**max_number_of_angular_momenta** The maximum number of angular momenta to be considered for non-local Kleinman-Bylander separable norm-conserving pseudopotentials. If there is no non-local part, set it to 0. If the s channel is the highest angular momentum channel over all atomic species, then set it to 1. If the p channel (resp. d or f) is the highest, set it to 2 (resp. 3 or 4).

**max_number_of_projectors** The maximum number of projectors for non-local Kleinman-Bylander separable norm-conserving pseudopotentials, over all angular momenta and all atomic species. If there is no non-local part, set it to 0. Most separable norm-conserving pseudopotentials have only one projector per angular momentum channel.

**kb_formfactor_sign** An array of integers whose value depend on the specific atomic species, angular momentum, and projector. It can have three values : when 0, it means that there is no projector defined for that channel. When +1 or -1, it gives the sign of the Kleinman-Bylander projector for that channel, as explained in section 5.2 .

**kb_formfactors**
**kb_formfactor_derivatives**
  Kleinman-Bylander form factors in reciprocal space, and their derivative, as explained in section 5.2 .

## 5.2. Comments

(1) On the density, kpoint weights and occupation numbers.

Supposing $\rho_{n,k}(r)$ to be the partial density at point r (in real space, using reduced coordinates) due to band n at k-point k (in reciprocal space, using reduced coordinates), then the full density at point is obtained thanks to

$$\rho(r_\alpha^{red}) = \sum_{s \in sym} \sum_k w_k \sum_n f_{n,k} \rho_{n,k}\left(S_{s,\alpha\beta}^{red}(r_\beta^{red} - t_{s,\beta}^{red})\right)$$

where $w_k$ is contained in the array "kpoint_weights" of Table 12, and

$f_{n,k}$ is contained in the array "occupations" of Table 13.

This relation generalizes to the collinear spin-polarized case, as well as the non-collinear case by taking into account the "number_of_components" defined in Table 5 , and the direction of the magnetization vector.

(2) On the Kleinman-Bylander form factors.

One can always write the non-local part of Kleinman-Bylander pseudopotential (reciprocal space) in the following way :

$$v_{nonloc}^{KB}(\vec{K}, \vec{K}') = \sum_s \left[\sum_{a(s)} e^{-i(\vec{K} - \vec{K}')\vec{\tau}_a}\right]\left[\sum_{lp} P_l(\hat{K}.\hat{K}')F_{slp}^*(K)S_{slp}F_{slp}(K')\right]$$

with $\vec{K} = \vec{k} + \vec{G}$ , $\vec{k}$ is one of the kpoints (see Table 12), $\vec{G}$ is a vector of the reciprocal lattice, the list of reduced coordinates of which can be found in the variable reduced_coordinates_of_plane_waves of Table 14. $K$ is the module of $\vec{K}$ and $\hat{K}$ its direction. $\vec{\tau}_a$ is the atomic position of atom $a$ belonging to species s. $P_l(x)$ is the Legendre polynomial of order $l$ . $F_{slp}(K)$ is the Kleinman-Bylander form factor for species s, angular momentum $l$, and number of projector p . $S_{slp}$ is the sign of the dyadic product $F^{*}_{slp}(K)F_{slp}(K')$. The sum on $a$ (s) runs over all atoms of atomic species s, $l$ runs over all the pseudopotential angular momentum components of the atomic species s, and p runs over the number of projectors allowed for a specific angular momentum channel of atomic species s. The additional variable kb_formfactor_derivative is equal to $\dfrac{dF_{slp}(K)}{dK}$ .

## 6. Concluding remarks

We presented the specifications for a file-format relying on the NetCDF I/O library with a content related to electronic structure and crystallographic data. This specification takes advantage of all the interesting properties of NetCDF-based files, in particular portability and extensibility. It is designed for both serial and distributed usage, although the latter characteristics was not presented here.

Several software in the Nanoquanta context can produce or read this file format : ABINIT [16,17], DP [18], GWST, SELF [19], V_Sim [20]. In order to further encourage its use, a library of Fortran routines [5] has been set up, and is available under the GNU LGPL licence.

## Acknowledgments

We thank the participants to the Louvain-la-neuve mini-workshop held in November 2005, at which a first instance of the density specification was outlined, especially P. Giannozzi, A. Marini, G. Bussi, A. Castro, M. Giantomassi, A. Incze, F. Jollet, J. J. Mortensen, M. Oliveira, S. Pesant, L. E. Ramos, G.-M. Rignanese, R. Shaltaf, F. Sottile, M. Stankovski.

We also acknowledge the help, interest and encouragements of several Nanoquanta members including R.W. Godby, L. Reining, A. Rubio, T. Patman, G. Bruant, C. Hogan. This work was carried out in the Nanoquanta EU Network of Excellence (NMP4-CT-2004-500198). M.J.V. acknowledges support from Marie Curie fellowship MANET (MEIF-CT-2005-024152).

 

| Attributes | Type (length) | Notes |
|---|---|---|
| file_format | char (80) | "ETSF Nanoquanta" |
| file_format_version | real | 2.1 or 2.2 or 3.0 ... |
| Conventions | char (80) | "http://www.etsf.eu/fileformats" |

Table 1. Mandatory global attributes for NQ/ETSF NetCDF files.

| Attributes | Type (length) | Notes |
|---|---|---|
| history | char (1024) | |
| title | char (80) | |

Table 2. Optional global attributes for NQ/ETSF NetCDF files.

| Attributes | Type (length) | Notes |
| --- | --- | --- |
| units | char (80) | required for variables that carry units |
| scale_to_atomic_units | double | required for units other than "atomic units" |

Table 3. Generic attributes that might be mandatory for selected variables in NQ/ETSF NetCDF files.

| Dimensions | Type | Notes |
|---|---|---|
| character_string_length | integer | Always ==80 |
| real_or_complex_coefficients | integer | Either ==1 or 2 |
| real_or_complex_density | integer | Either ==1 or 2 |
| real_or_complex_gw_corrections | integer | Either ==1 or 2 |
| real_or_complex_potential | integer | Either ==1 or 2 |
| real_or_complex_wavefunctions | integer | Either ==1 or 2 |
| number_of_cartesian_directions | integer | Always ==3 |
| number_of_reduced_dimensions | integer | Always ==3 |
| number_of_vectors | integer | Always ==3 |
| number_of_symmetry_operations | integer | |
| number_of_atoms | integer | |
| number_of_atom_species | integer | |
| symbol_length | integer | Always ==2 |

Table 4. Variables that specify dimensions in NQ/ETSF NetCDF files (no splitting case see Ref.[4]).

| Dimensions | Type | Notes |
|---|---|---|
| max_number_of_states | integer | |
| number_of_kpoints | integer | |
| number_of_spins | integer | Either == 1 or 2 |
| number_of_spinor_components | integer | Either == 1 or 2 |
| number_of_components | integer | Either == 1, 2 or 4 |
| max_number_of_coefficients | integer | |
| number_of_grid_points_vector1 | integer | |
| number_of_grid_points_vector2 | integer | |
| number_of_grid_points_vector3 | integer | |

Table 5. Variables that specify dimensions in NQ/ETSF NetCDF files (possible splitting case, see Ref.[4]).

| Variables | Type (index order as in C) | Notes |
|---|---|---|
| valence_charges | double[number_of_atom_species] | |
| pseudopotential_types | char[number_of_atom_species] [character_string_length] | |

Table 6. Optional variables : atomic information.

| Variables | Type (index order as in C) | Notes |
|---|---|---|
| number_of_electrons | integer | |
| exchange_functional | char [character_string_length] | |
| correlation_functional | char [character_string_length] | |
| fermi_energy | double | Units attribute required. The attribute "scale_to_atomic_units" might also be mandatory, see section 2.2. |
| smearing_scheme | char [character_string_length] | |
| smearing_width | double | Units attribute required. The attribute "scale_to_atomic_units" might also be mandatory, see section 2.2. |

Table 7. Optional variables : electronic structure.

| Variables | Type (index order as in C) | Notes |
|---|---|---|
| kinetic_energy_cutoff | double | Units attribute required. The attribute "scale_to_atomic_units" might also be mandatory, see section 2.2. |
| kpoint_grid_shift | double [number_of_reduced_dimensions] | |
| kpoint_grid_vectors | double[number_of_vectors] [number_of_reduced_dimensions] | |
| monkhorst_pack_folding | integer [number_of_vectors] | |

Table 8. Optional variables : reciprocal space.

| Variables | Type (index order as in C) | Notes |
|---|---|---|
| primitive_vectors | double[number_of_vectors] [number_of_cartesian_directions] | By default, given in Bohr |
| reduced_symmetry_matrices | integer[number_of_symmetry_operations] [number_of_reduced_dimensions] [number_of_reduced_dimensions] | The "symmorphic" attribute is needed. |
| reduced_symmetry_translations | double[number_of_symmetry_operations] [number_of_reduced_dimensions] | The "symmorphic" attribute is needed. |
| space_group | integer | Between 1 and 232 |
| atom_species | integer[number_of_atoms] | Between 1 and "number_of_atom_species" |
| reduced_atom_positions | double[number_of_atoms] [number_of_reduced_dimensions] | |
| atomic_numbers | double[number_of_atom_species] | |
| atom_species_names | char[number_of_atom_species] [character_string_length] | |
| chemical_symbols | char[number_of_atom_species] [symbol_length] | |
| Attributes | Type | |
| symmorphic | char(80) | flag-type attribute, see section 2.3 |

Table 9. Variables and attributes to specify the atomic structure and symmetry operations.

| Variables | Type (index order as in C) | Notes |
|---|---|---|
| density | double[number_of_components]<br>[number_of_grid_points_vector3]<br>[number_of_grid_points_vector2]<br>[number_of_grid_points_vector1]<br>[real_or_complex_density] | By default, the density is given in atomic units, that is, number of electrons per Bohr^3<br>The "units" attribute is required.<br>The attribute "scale_to_atomic_units" might also be mandatory, see section 2.2. |

Table 10. The specification of the density (the last of the variables).

| Variables | Type (index order as in C) | Notes |
|---|---|---|
| correlation_potential | double[number_of_components] | Units attribute required. |
| and/or | [number_of_grid_points_vector3] | The attribute |
| exchange_potential | [number_of_grid_points_vector2] | "scale_to_atomic_units" |
| and/or | [number_of_grid_points_vector1] | might also be mandatory, see |
| exchange_correlation_potential | [real_or_complex_potential] | section 2.2. |

Table 11. The specification of exchange, correlation, and exchange-correlation potentials.

| Variables | Type (index order as in C) | Notes |
|:---:|:---|:---:|
| reduced_coordinates_of_kpoints | double[number_of_kpoints] [number_of_reduced_dimensions] | |
| kpoint_weights | double[number_of_kpoints] | See description of density construction in section 5.2 |

Table 12. Variables that specify the k points.

| Variables | Type (index order as in C) | Notes |
|---|---|---|
| number_of_states | integer[number_of_spins] [number_of_kpoints] | The attribute "k_dependent" must be defined |
| eigenvalues | double[number_of_spins] [number_of_kpoints] [max_number_of_states] | The "units" attribute is required. The attribute "scale_to_atomic_units" might also be mandatory, see section 2.2. See also possibles changes for split files in Table 18 |
| occupations | double[number_of_spins] [number_of_kpoints] [max_number_of_states] | See also possibles changes for split files in Table 18 |
| Attributes | Type (index order as in C) | |
| k_dependent | char(80) | Attribute to number_of_states, flag-type, see section 2.3 |

Table 13. Specifications related to each state : occupation numbers and eigenvalues.

| Variables | Type (index order as in C) | Notes |
|---|---|---|
| basis_set | char(character_string_length) | "plane_waves" if a plane-wave basis set is used |
| number_of_coefficients | integer[number_of_kpoints] | The attribute "k_dependent" must be defined (see Table 15). Possible splitting, see Table 18. |
| reduced_coordinates_of_plane_waves | integer[number_of_kpoints] [max_number_of_coefficients] [number_of_reduced_dimensions] | The attribute "k_dependent" must be defined (see Table 15). The attribute used_time_reversal_at_gamma might be defined. |
| coefficients_of_wavefunctions | double [number_of_spins] [number_of_kpoints] [max_number_of_states] [number_of_spinor_components] [max_number_of_coefficients] [real_or_complex_coefficients] | Normalization : 1 per unit cell, see section 5.2 The attribute used_time_reversal_at_gamma might be defined. |
| Attributes | Type (index order as in C) | |
| used_time_reversal_at_gamma | char(80) | Attribute to reduced_coordinaters_of_plane_waves and coefficients_of_wavefunctions flag-type, see section 2.3 |

Table 14 Specification of wavefunctions in a plane-wave basis. Needed only in case "coefficients_of_wavefunctions" will be the array containing the wavefunctions.

| Variables | Type (index order as in C) | Notes |
| --- | --- | --- |
| real_space_wavefunctions | double [number_of_spins] [number_of_kpoints] [max_number_of_states] [number_of_spinor_components] [number_of_grid_points_vector3] [number_of_grid_points_vector2] [number_of_grid_points_vector1] [real_or_complex_wavefunctions] | Normalization : 1 per unit cell. See possible modifications for split files in Table 18 |

Table 15. Specification of wavefunctions in real space.

| Dimensions | Type (index order as in C) | Notes |
|---|---|---|
| max_number_of_angular_momenta | integer | |
| max_number_of_projectors | integer | |

| Variables | Type (index order as in C) | Notes |
|---|---|---|
| gw_corrections | double[number_of_spins]<br>[number_of_kpoints]<br>[max_number_of_states]<br>[real_or_complex_gw_corrections] | The "units" attribute is required.<br>The attribute "scale_to_atomic_units" might also be mandatory, see section 2.2. |
| kb_formfactor_sign | integer[number_of_atom_species]<br>[max_number_of_angular_momenta]<br>[max_number_of_projectors] | |
| kb_formfactors | double[number_of_atom_species]<br>[max_number_of_angular_momenta]<br>[max_number_of_projectors]<br>[number_of_kpoints]<br>[max_number_of_coefficients] | |
| kb_formfactor_derivative | double[number_of_atom_species]<br>[max_number_of_angular_momenta]<br>[max_number_of_projectors]<br>[number_of_kpoints]<br>[max_number_of_coefficients] | |

Table 16. Optional dimensions and variables that might be needed for some GW/BSE softwares